# An $O(bn^2)$ Time Algorithm for Optimal Buffer Insertion with $b$ Buffer Types *

Zhuo Li
Dept. of Electrical Engineering
Texas A&M University
College Station, Texas 77843, USA.
zhuoli@ee.tamu.edu

Weiping Shi
Dept. of Electrical Engineering
Texas A&M University
College Station, Texas 77843, USA.
wshi@ee.tamu.edu

**Abstract**

*Buffer insertion is a popular technique to reduce the interconnect delay. The classic buffer insertion algorithm of van Ginneken has time complexity $O(n^2)$, where $n$ is the number of buffer positions. Lillis, Cheng and Lin extended van Ginneken's algorithm to allow $b$ buffer types in time $O(b^2n^2)$. For modern design libraries that contain hundreds of buffers, it is a serious challenge to balance the speed and performance of the buffer insertion algorithm.*

*In this paper, we present a new algorithm that computes the optimal buffer insertion in $O(bn^2)$ time. The reduction is achieved by the observation that the $(Q,C)$ pairs of the candidates that generate the new candidates must form a convex hull. On industrial test cases, the new algorithm is faster than the previous best buffer insertion algorithms by orders of magnitude.*

## 1. Introduction

Delay optimization techniques for interconnect are increasingly important for achieving timing closure of high performance designs. One popular technique for reducing interconnect delay is buffer insertion. A recent study by Saxena *et al.* [10] projects that 35% of all cells will be intra-block repeaters for the 45 nm node. Consequently, algorithms that can efficiently insert buffers are essential for the design automation tools.

In 1990, van Ginneken [14] proposed an optimal buffer insertion algorithm for one buffer type. His algorithm has time complexity $O(n^2)$, where $n$ is the number of candidate buffer positions. Lillis, Cheng and Lin [7] extended van Ginneken's algorithm to allow $b$ buffer types in time $O(b^2n^2)$. Recently, Shi and Li [11] presented a new algorithm with time complexity $O(n\log n)$ for 2-pin nets, and $O(n\log^2 n)$ for multi-pin nets, for one buffer type. Several works have built upon van Ginneken's algorithm and its extension for multiple buffer types to include wire sizing [7], simultaneous tree construction [8, 6, 5, 9, 15], noise constraints [2] and resource minimization [7, 13].

Modern design libraries may contain hundreds of different buffers with different input capacitances, driving resistance, intrinsic delay, power level, etc. If every buffer available for the given technology is supplied, it is stated in [3] that the current algorithms could possibly take days or even weeks for large designs since all these algorithms are quadratic in terms of $b$. Alpert et. al. [3] studied how to reduce the size of the buffer library with a clustering algorithm. Though the buffer library size is reduced, the solution quality is often degraded accordingly.

In this paper, we propose a new algorithm that performs optimal buffer insertion with $b$ buffer types in $O(bn^2)$ time. Our speedup is achieved by the observation that the candidates that generate new buffered candidates must lie on the convex hull of $(Q,C)$. Experimental results show that our algorithm is significantly faster than previous best algorithms.

Section 2 formulates the problem. Section 3 describes the new algorithm. Simulation results are given in Section 4 and conclusions are given in Section 5.

## 2. Preliminary

A net is given as a routing tree $T = (V,E)$, where $V = \{s_0\} \cup V_s \cup V_n$, and $E \subseteq V \times V$. Vertex $s_0$ is the *source* vertex and also the root of $T$, $V_s$ is the set of *sink* vertices, and $V_n$ is the set of *internal* vertices. Each sink vertex $s \in V_s$ is associated with sink capacitance $C(s)$ and required arrival time $RAT(s)$. A buffer library $B$ contains different types of buffers and its size is represented by $b$. For each buffer type $B_i \in B$, the intrinsic delay is $K(B_i)$, driving resistance is $R(B_i)$, and input capacitance is $C(B_i)$. A function $f : V_n \to 2^B$ specifies the types of buffers allowed at each internal vertex. Each edge $e \in E$ is associated with lumped resistance $R(e)$ and capacitance $C(e)$.

---

* This research was supported by the NSF grants CCR-0098329, CCR-0113668, EIA-0223785, ATP grant 512-0266-2001.

Following previous researchers [14, 7, 9, 15, 1], we use the Elmore delay for the interconnect and the linear delay for buffers. For each edge $e = (v_i, v_j)$, signals travel from $v_i$ to $v_j$. The Elmore delay of $e$ is

$$D(e) = R(e)\left(\frac{C(e)}{2} + C(v_j)\right),$$

where $C(v_j)$ is the downstream capacitance at $v_j$. For any buffer type $B_i$ at vertex $v_j$, the buffer delay is

$$D(v_j) = R(B_i) \cdot C(v_j) + K(B_i),$$

where $C(v_j)$ is the downstream capacitance at $v_j$. When a buffer $B_i$ is inserted, the capacitance viewed from the upper stream is $C(B_i)$.

For any vertex $v \in V$, let $T(v)$ be the subtree downstream from $v$, and with $v$ being the root. Once we decide where to insert buffers in $T(v)$, we have a **candidate** $a$ for $T(v)$. The delay from $v$ to sink $s \in T(v)$ under $a$ is

$$D(v, s, \alpha) = \sum_{e=(v_i, v_j)} (D(v_i) + D(e)),$$

where the sum is over all edges $e$ in the path from $v$ to $s$. If $v_i$ is a buffer in $\alpha$, then $D(v_i)$ is the buffer delay. If $v_i$ is not a buffer in $a$, then $D(v_i) = 0$. The slack of $v$ under $a$ is

$$Q(v, a) = \min_{s \in T(v)} \{RAT(s) - D(v, s, a)\}.$$

**Buffer Insertion Problem:** Given routing tree $T = (V, E)$, sink capacitance $C(s)$ and $RAT(s)$ for each sink $s$, capacitance $C(e)$ and resistance $R(e)$ for each edge $e$, possible buffer position $f$, and buffer library $B$, find a candidate $a$ for $T$ that maximizes $Q(s_0, a)$.

The effect of a candidate to the upstream is described by slack $Q$ and downstream capacitance $C$ [14]. Define $C(v, \alpha)$ as the downstream capacitance at node $v$ under candidate $a$. For any two candidates $\alpha_1$ and $\alpha_2$ of $T(v)$, we say $\alpha_1$ **dominates** $\alpha_2$, if $Q(v, \alpha_1) \geq Q(v, \alpha_2)$ and $C(v, \alpha_1) \leq C(v, \alpha_2)$. The set of **nonredundant candidates** of $T(v)$, which we denote as $N(v)$, is the set of candidates such that no candidate in $N(v)$ dominates any other candidate in $N(v)$, and every candidate of $T(v)$ is dominated by some candidates in $N(v)$. Once we have $N(s_0)$, the candidate that gives the maximum $Q(s_0, a)$ can be found easily. The number of total nonredundant candidates is at most $n + 1$ for one buffer type and $bn + 1$ for $b$ buffer types [14, 7], where n is the number of candidate buffer positions.

## 3. New Algorithm

The previous best algorithm for multiple buffer types by Lillis, Cheng and Lin consists of three major operations: 1) adding buffers at a buffer position in $O(b^2 n)$ time, 2) adding a wire in $O(bn)$ time, and 3) merging two branches in $O(bn_1 + bn_2)$ time, where $n_1$ and $n_2$ are the numbers of buffer positions in the two branches. As a result, their algorithm has time complexity $O(b^2 n^2)$. In this section, we show that the time complexity of the first operation, adding buffers at a buffer position, can be reduced to $O(bn)$, and thus our algorithm can achieve total time complexity $O(bn^2)$.

Assume we have computed the set of nonredundant candidates $N(v_1)$ for $T(v_1)$, and now reach a buffer position $v$, see Fig. 1. Wire $(v, v_1)$ has $0$ resistance and capacitance. Define $P_i(\alpha)$ as the slack if we add a buffer type $B_i$ at v for any candidate $a$ in $N(v_1)$:

$$P_i(\alpha) = Q(v_1, \alpha) - R(B_i) \cdot C(v_1, \alpha) - K(B_i). \quad (1)$$

If we do not insert any buffer at $v$, then every candidate for $T(v_1)$ is a candidate for $T(v)$. If we insert a buffer at $v$, then for every buffer type $B_i$, $i = 1, 2, \ldots, b$, there will be a new candidate $\beta_i$:

$$Q(v, \beta_i) = \max_{\alpha \in N(v_1)} \{P_i(\alpha)\},$$
$$C(v, \beta_i) = C(B_i).$$

Note that some of the new candidates $\beta_i$ could be redundant. The algorithm of Lillis, Cheng and Lin takes $O(b^2 n)$ time to generate all $\beta_i$s and $O(b^2 n)$ time to insert nonredundant ones into the list of nonredundant candidates $N(v_1)$.

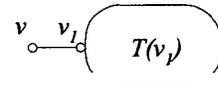

**Figure 1.** $T(v)$ **consists of buffer position** $v$ **and** $T(v_1)$.

We show how to generate all $\beta_i$s in $O(bn)$ time. Since all candidates discussed in this section are in $N(v_1)$, we will write $Q(\alpha)$ for $Q(v_1, a)$, and $C(\alpha)$ for $C(v_1, a)$. Suppose buffers in the buffer library are sorted according to its driving resistance $R(B_i)$ in non-increasing order, $R(B_1) \geq R(B_2) \geq \ldots \geq R(B_b)$. If some buffer types are not allowed at $v$, we simply omit them without affecting the rest of the algorithm. **For any buffer type $B_i \in B$, define the best candidate** for $B_i$ as the candidate $\alpha_i \in N(v_1)$ such that $\alpha_i$ maximizes $P_i(\alpha)$ among all candidates of $N(v_1)$. If there are multiple $a$'s that maximize $P_i(a)$, the one with minimum $C(\alpha)$ is chosen.

**Lemma 1** For any two buffer types $B_i$ and $B_j$, where $i > j$, let their best candidates be $\alpha_i$ and $\alpha_j$, respectively. Then we must have $C(\alpha_i) \geq C(\alpha_j)$.

**Proof:** From the definition of $\alpha_i$, we have $P_i(\alpha_i) \geq P_i(\alpha_j)$ and $P_j(\alpha_j) \geq P_j(\alpha_i)$. Consequently,

$$Q(\alpha_i) - Q(\alpha_j) \geq R(B_i) \cdot (C(\alpha_i) - C(\alpha_j)),$$
$$Q(\alpha_j) - Q(\alpha_i) \geq R(B_j) \cdot (C(\alpha_j) - C(\alpha_i)).$$

Therefore, $(R(B_i) - R(B_j))(C(\alpha_i) - C(\alpha_j)) \leq 0$.

Since $i > j$, $R(B_j) \geq R(B_i)$. If $R(B_j) > R(B_i)$, $C(\alpha_i) \geq C(\alpha_j)$. If $R(B_j) = R(B_i)$, then it is easy to get $P_i(\alpha_i) = P_i(\alpha_j)$ and $P_j(\alpha_j) = P_j(\alpha_i)$. From the definition, when there are multiple $a$'s that maximize $P_i(\alpha)$, the one with minimum $C(\alpha)$ is chosen. Thus $\alpha_i$ and $\alpha_j$ should be the same candidate, which means $C(\alpha_i) = C(\alpha_j)$. □

Lemma 1 implies that the best candidates $\alpha_1, \ldots, \alpha_b$ for buffer types $B_1, \ldots, B_b$ are in increasing order of $C$. However, this is not enough for an $O(bn^2)$ time algorithm. In the following, we define the concept of **convex pruning**, which is important in generating new candidates $\beta_i$'s.

**Convex pruning:** Let $\alpha_1, \alpha_2$ and $\alpha_3$ be three nonredundant candidates of $T(v_1)$ such that $C(\alpha_1) < C(\alpha_2) < C(\alpha_3)$. If

$$\frac{Q(\alpha_2) - Q(\alpha_1)}{C(\alpha_2) - C(\alpha_1)} < \frac{Q(\alpha_3) - Q(\alpha_2)}{C(\alpha_3) - C(\alpha_2)}, \quad (2)$$

then we prune candidate $\alpha_2$.

Convex pruning can be explained by Figure 2. Consider $Q$ as the Y-axis and $C$ as the X-axis. Then the set of nonredundant candidate $N(v_1)$ are a set of points in the two-dimensional plane. Candidate $\alpha_2$ in the above definition is shown in Figure 2(a), and is pruned in Figure 2(b). Call the candidates after convex pruning $M(v_1)$. It can be seen that $N(v_1)$ is a monotonically increasing sequence, while $M(v_1)$ is a convex hull.

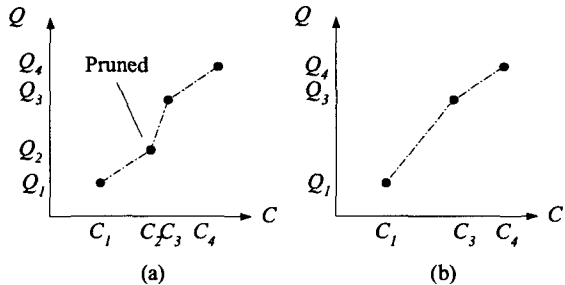

**Figure 2.** (a) Nonredundant candidates $N(v_1)$ on $(Q, C)$ plane. (b) Nonredundant candidates $M(v_1)$ after convex pruning.

Function **Convexpruning** performs convex pruning for any list of nonredundant candidates sorted in increasing $Q$ and $C$ order. The following C code defines the data structure for each candidate in the list:

```
typedef struct Candidate {
    float Q, C;
    struct Candidate *next, *prev;
        // double link list
} Candidate;
```

Let the candidate with minimum $C$ be $\alpha_1$. We add a dummy candidate $(-\infty, C(\alpha_1))$ pointed by **header** to simplify the algorithm. Function **LeftTurn** checks if **a1**, **a2** and **a3** form a left turn on the plane. It is the same as the condition in Eq. (2).

```
void ConvexPruning(Candidate *header)
{
    Candidate *a1, *a2, *a3;

    a1 = header;
    a2 = a1->next;
    a3 = a2->next;

    while (a3 != NULL) {
        if (LeftTurn(a1, a2, a3)) {
            // prune a2 and move backward
            free(a2);
            a1->next = a3;
            a3->prev = a1;
            a2 = a1;
            a1 = a1->prev;
        } else {
            // move forward
            a3 = a3->next;
            a2 = a2->next;
            a1 = a1->next;
        }
    }
}
```

**Lemma 2** Given any set of $k$ nonredundant candidates sorted in increasing $Q$ and $C$ order, function ConvexPruning performs convex pruning in $O(k)$ time.

**Proof:** This procedure is known as **Graham's** scan in computational geometry [4]. It finds the convex hull of a set of points in sorted order in linear time.

It is well known that a set of points form a convex hull if and only if there are no consecutive $\alpha_1, \alpha_2$ and $\alpha_3$ that satisfy Eq. (2). Therefore, **Convexpruning** is correct since it checks all consecutive candidates.

To analyze the time complexity, consider the number of forward and backward moves. Each time

`ConvexPruning` moves backward, it deletes a candidate. Therefore, there can be at most $k$ backward moves. The number of forward moves is the size of the list plus the number of backward moves. Therefore the number of forward moves is at most $2k$. Hence the time complexity is $O(k)$. □

**Lemma 3** *For any buffer type $B_i \in \boldsymbol{B}$, its best candidate $\alpha_i$ that maximizes $P_i(\alpha)$ is not pruned by* `ConvexPruning`.

**Proof:** Consider any candidate $\gamma \in N(v_1)$ with $C(\gamma) > C(\alpha_i)$. According to the definition of $\alpha_i$, we have $P_i(\alpha_i) \geq P_i(\gamma)$. Therefore,

$$Q(\gamma) - Q(\alpha_i) \leq R(B_i) \cdot (C(\gamma) - C(\alpha_i)),$$
$$\frac{Q(\gamma) - Q(\alpha_i)}{C(\gamma) - C(\alpha_i)} \leq R(B_i).$$

Similarly for any candidate $\eta \in N(v_1)$ with $C(\eta) < C(\alpha_i)$, we have

$$Q(\alpha_i) - Q(\eta) \geq R(B_i) \cdot (C(\alpha_i) - C(\eta)),$$
$$\frac{Q(\alpha_i) - Q(\eta)}{C(\alpha_i) - C(\eta)} \geq R(B_i).$$

Therefore,

$$\frac{Q(\alpha_i) - Q(\eta)}{C(\alpha_i) - C(\eta)} \geq \frac{Q(\gamma) - Q(\alpha_i)}{C(\gamma) - C(\alpha_i)},$$

where $\eta$ is any candidates with $C(\eta) < C(\alpha_i)$, and $\gamma$ is any candidates with $C(\gamma) > C(\alpha_i)$. According to the definition of convex pruning, $\alpha_i$ is not pruned. □

**Lemma 4** *Let the set of nonredundant candidates after* `Convexpruning` *be $M(v_1)$ and assume $M(v_1)$ are sorted in increasing $Q$ and $C$ order. Consider any three candidates $\eta, \alpha, \gamma$ in $M(v_1)$, such that $C(\eta) < C(\alpha) < C(\gamma)$. For any buffer type $B_i \in \boldsymbol{B}$, if $P_i(\eta) \geq P_i(\alpha)$, then $P_i(\eta) \geq P_i(\gamma)$; if $P_i(\gamma) \geq P_i(\alpha)$, then $P_i(\gamma) \geq P_i(\eta)$.*

**Proof:** From the definition of convex pruning, we have

$$\frac{Q(\gamma) - Q(\alpha)}{C(\gamma) - C(\alpha)} \leq \frac{Q(\alpha) - Q(\eta)}{C(\alpha) - C(\eta)}.$$

If $P_i(\eta) \geq P_i(\alpha)$, then

$$\frac{Q(\alpha) - Q(\eta)}{C(\alpha) - C(\eta)} \leq R(B_i),$$
$$\frac{Q(\gamma) - Q(\alpha)}{C(\gamma) - C(\alpha)} \leq R(B_i),$$
$$Q(\alpha) - R(B_i) \cdot C(\alpha) \geq Q(\gamma) - R(B_i) \cdot C(\gamma),$$
$$P_i(\alpha) \geq P_i(\gamma).$$

Therefore, $P_i(\eta) \geq P_i(\gamma)$. Similarly, if $P_i(\gamma) \geq P_i(\alpha)$, then

$$\frac{Q(\gamma) - Q(\alpha)}{C(\gamma) - C(\alpha)} \geq R(B_i),$$
$$\frac{Q(\alpha) - Q(\eta)}{C(\alpha) - C(\eta)} \geq R(B_i),$$
$$Q(\alpha) - R(B_i) \cdot C(\alpha) \geq Q(\eta) - R(B_i) \cdot C(\eta)$$
$$P_i(\alpha) \geq P_i(\eta).$$

Therefore, $P_i(\gamma) \geq P_i(\eta)$. □

Lemma **4** implies that for any buffer type $B_i$, if candidate $a$ maximizes $P_i(\alpha)$ among its previous and next consecutive candidates in $M(v_1)$, then $a$ maximizes $P_i(\alpha)$ among all candidates in $M(v_1)$.

Function `AddBuffer` identifies $\alpha_i$ from $N(v_1)$ and generates new candidates $\beta_i, i = 1,\ldots,b$. Nonredundant candidates in $N(v_1)$ are stored in increasing $C$ order using a double link list pointed by `header`. Buffer types are sorted in non-increasing driver resistance order and stored in array B. Function `P(i, a)` computes $P_i(\alpha)$ as defined in **Eq.** (1).

```
Candidate *AddBuffer(Candidate *header)
{
    Candidate *a1, *a2;
    Candidate *beta[];
    int i;

    Convexpruning (header);

    a1 = header;
    a2 = a1->next;
    for (i = 1; i <= BUF_LIB_SIZE; i ++) {
        while (a2 != NULL) {
            if (P(i, a1) c P(i, a2)) {
                a1 = a1->next;
                a2 = a1->next;
            } else
                break;
        }

        // generate new candidate beta_i
        beta[i]->Q = P(i, a1);
        beta[i]->C = B[i]->C;
    }

    sort beta's in nondecreasing C order;
    return beta's;
}
```

**Theorem 1** *If $v$ is a buffer position, wire $(v, v_1)$ is a wire with zero resistance and capacitance, nonredundant candidates of $N(v_1)$ are stored in increasing $Q$ and $C$ order, then* *function* `AddBuffer` *generates all new candidates* $\beta_i$ *from* $N(v_1)$ *in* $O(bn)$ *time.*

**Proof:** Let the set of nonredundant candidates after `Convexpruning` be $M(v_1)$. From Lemma 3, we know that all best candidates $\alpha_i$'s are in $M(v_1)$. From Lemma 1 and Lemma 4, starting from the first candidates in $M(v_1)$, function `AddBuffer` can find all $\beta_i$s in the increasing order of $i$.

Now consider the time complexity. Function `Convexpruning` takes $O(bn)$ time according to Lemma 2. The `for` loop takes $O(bn + b) = O(bn)$ time. It takes only $O(b \log b)$ time to sort the entire buffer library in terms of the input capacitance $C(B_i)$, and establish the order from buffer index $i$ to the order in $C(B_i)$. Each time function `AddBuffer` is called, the new candidates $\beta_i$'s can be sorted in nondecreasing $C$ order by using the index in $O(b)$ time. □

**Theorem 2** *Given a set of $O(bn)$ nonredundant candidates sorted in increasing $Q$ and $C$ order, all $b$ new candidates $\beta_i$'s can be inserted in $O(bn)$ time.*

**Proof:** Since $\beta_i$'s are in the nondecreasing order of capacitance $C(\beta_i)$ and the given set of nonredundant candidates are in nondecreasing order of $C(\alpha)$, it takes $O(bn) + O(b) = O(bn)$ time to merge the two sorted lists. □

Since the operation of adding a buffer is reduced to $O(bn)$ time from Theorem 1 and 2, it is easy to see that buffer insertion with $b$ buffer types can be done in worst case time $O(bn^2)$ with our new algorithm.

## 4. Simulation

Both the algorithm of Lillis et al. [7] and the new algorithm are implemented in C and run on a Sun SPARC workstations with 400 MHz and 2 GB memory. The device and interconnect parameters are based on TSMC 180 nm technology. We have 4 different buffer libraries, with the size 8, 16, 32 and 64 respectively. $R(B_i)$ is chosen from 180 $\Omega$ to 7000 $\Omega$, $C(B_i)$ is chosen from 0.7 fF to 23 fF, and $K(B_i)$ is chosen from 29 ps to 36.4 ps. The sink capacitances range from 2 fF to 41 fF. The wire resistance is 0.076 $\Omega/\mu m$ and the wire capacitance is 0.118 fF/$\mu m$. Table 1 shows for large industrial circuits, the new algorithm is up to 11 times faster than Lillis' algorithm. The memory usage is not shown in the table, but there is only almost 2% memory overhead due to the double linked list used by the new algorithm. When $b$ is small, $O(bn^2)$ algorithm has a little time overhead compared to Lillis' algorithm. due to function `ConvexPruning`.

Fig. 3 shows the time complexity curve of two algorithms for the net with 1944 sinks and 33133 buffer positions with respect to the size of buffer library $b$. In the figure, the y axis is normalized to the running time of the case when the buffer library size is 8. Though the worst case time complexity of Lillis' algorithm is quadratic in terms of $b$, it behaves more like a linear function of $b$, as observed in [3]. The time complexity curve of our algorithm is also linear, but has a much smaller slope.

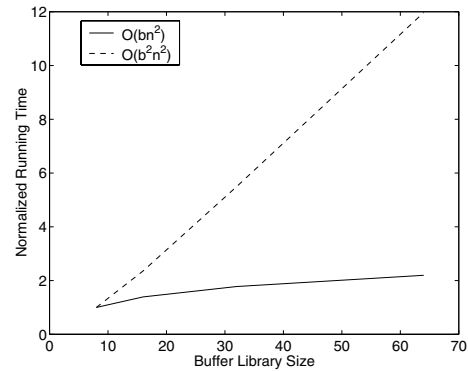

**Figure 3. Comparison of normalized running time with respect to buffer library size among two algorithms. Number of sink is 1944 and number of buffer positions is 33133.**

Fig. 4 shows the time complexity curve of the two algorithms for the net with 1944 sinks, with respect to the number of buffer positions $n$. The buffer library size is 32. In the figure, the y axis is normalized to the running time of the case with 1943 buffer positions. We can see that while Lillis' and our algorithms both behave quadratically, our algorithm shows much slower growing trend since the operation of adding a buffer becomes more dominant among three major operations when n increases.

## 5. Conclusion

We presented a new algorithm for optimal buffer insertion with $b$ buffer types of worst case time $O(bn^2)$. This is an improvement of the previous best $O(b^2n^2)$ algorithm[7]. Simulation results show our new algorithm is significantly faster than $O(b^2n^2)$ algorithms for large industrial circuits with large buffer libraries. Our algorithm can also be applied to reduce buffer cost. We leave the details to the journal version.

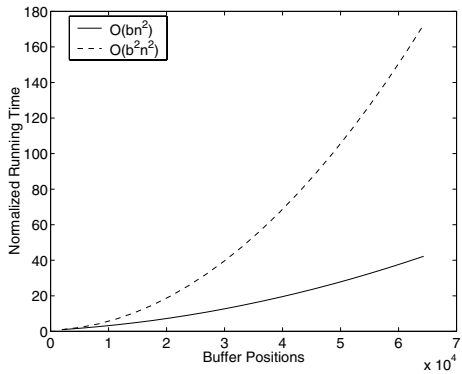

**Figure 4. Comparison of normalized running time with respect to buffer positions among two algorithms. Number of sink is 1944 and number of buffer types is 32.**

| $m$ | $n$ | $b$ | CPU Time (sec) | | Speed -up |
|---|---|---|---|---|---|
| | | | $O(bn^2)$ | $O(b^2n^2)$ [7] | |
| 337 | 336 | 8 | 0.08 | 0.09 | 1.11 |
| | | 16 | 0.14 | 0.16 | 1.14 |
| | | 32 | 0.23 | 0.36 | 1.57 |
| | | 64 | 0.42 | 0.91 | 2.17 |
| | 5647 | 8 | 1.54 | 2.15 | 1.40 |
| | | 16 | 2.11 | 4.55 | 2.16 |
| | | 32 | 2.81 | 9.99 | 3.56 |
| | | 64 | 4.05 | 22.52 | 5.56 |
| | 10957 | 8 | 4.56 | 7.15 | 1.57 |
| | | 16 | 6.02 | 15.74 | 2.61 |
| | | 32 | 7.62 | 34.02 | 4.46 |
| | | 64 | 9.98 | 74.55 | 7.47 |
| 1944 | 1943 | 8 | 0.93 | 0.90 | 0.97 |
| | | 16 | 1.62 | 1.86 | 1.15 |
| | | 32 | 2.78 | 4.38 | 1.58 |
| | | 64 | 4.54 | 10.71 | 2.36 |
| | 33133 | 8 | 22.96 | 38.19 | 1.66 |
| | | 16 | 31.97 | 90.08 | 2.82 |
| | | 32 | 40.83 | 209.82 | 5.14 |
| | | 64 | 50.42 | 457.22 | 9.07 |
| | 64323 | 8 | 70.23 | 141.07 | 2.01 |
| | | 16 | 95.78 | 337.97 | 3.53 |
| | | 32 | 117.38 | 755.46 | 6.44 |
| | | 64 | 136.85 | 1596.61 | 11.67 |
| 2676 | 2675 | 8 | 1.16 | 1.13 | 0.97 |
| | | 16 | 2.07 | 2.38 | 1.15 |
| | | 32 | 3.83 | 5.78 | 1.51 |
| | | 64 | 6.18 | 14.15 | 2.30 |
| | 45075 | 8 | 27.31 | 44.29 | 1.62 |
| | | 16 | 36.75 | 98.31 | 2.68 |
| | | 32 | 47.8 | 226.25 | 4.73 |
| | | 64 | 64.02 | 543.45 | 8.49 |
| | 87475 | 8 | 82.67 | 163.87 | 1.98 |
| | | 16 | 108.16 | 372.22 | 3.44 |
| | | 32 | 134.83 | 835.04 | 6.19 |
| | | 64 | 164.08 | 1864.08 | 11.36 |

**Table 1. Simulation results for industrial test cases, where $m$ is the number of sinks, $n$ is the number of buffer positions, and $b$ is the library size.**